\documentclass[preprint,epsf]{article}
\usepackage{epsf}
\usepackage[T1]{fontenc}
\begin{document}

\title{Hysteresis in an Electrochemical System:\\
Br Electrodeposition on Ag(100)}

\author{
\centerline{S.J. Mitchell, G. Brown, P.A. Rikvold}\\
\centerline{\footnotesize\it Center for Materials Research and Technology,}\\
\centerline{\footnotesize\it Department of Physics, and}\\
\centerline{\footnotesize\it School of Computational Science and 
Information Technology,}\\
\centerline{\footnotesize\it Florida State University, 
Tallahassee, Florida 32306-4351, USA}\\
}

\maketitle

\begin{abstract}
We examine hysteresis in cyclic voltammetry
of Br electrosorption onto single-crystal Ag(100)
by dynamic Monte Carlo simulations.
At room temperature, this system displays a second-order 
phase transition from a low-coverage disordered phase
to a doubly degenerate ${\rm c}(2 \times 2)$ phase.
The electrochemical potential is ramped back and forth
across the phase transition, linearly in time,
and the phase lag between the response of the adlayer and the
potential is observed to depend on the sweep rate.
The hysteresis in this system is caused by slow
ordering/disordering kinetics and critical slowing-down.
\end{abstract}

Hysteresis is a widely occurring phenomenon in which
a system is driven too fast to remain near equilibrium,
resulting in a phase lag between the response and the oscillating driving force.
Some common examples include 
magnetic systems in oscillating external fields \cite{SIDE99,X}, 
electrochemical adsorption in an oscillating electrochemical potential 
\cite{OCKO:BR/AG,ME:ECCHAPT,ME:BR2AG}, 
and nonlinear media under oscillating stress.
As an example of hysteresis at a second-order phase transition, we here present
preliminary results of dynamic Monte Carlo simulations of the electrosorption 
of Br onto single-crystal Ag(100)
under a time-varying electrochemical potential,
known as cyclic voltammetry (CV) in the electrochemical literature.

In equilibrium at room temperature, this system displays a second-order phase transition 
in the Ising universality class
from a low-coverage disordered phase at more negative potentials
to a doubly degenerate ${\rm c}(2 \times 2)$ ordered phase at more 
positive potentials \cite{OCKO:BR/AG}.
As the system is driven back and forth across this phase transition,
hysteresis is observed \cite{OCKO:BR/AG}.
Unlike magnetic systems \cite{SIDE99,X} and some electrochemical systems \cite{ME:ECCHAPT}, 
in which hysteresis is usually associated with
metastable decay near a first-order phase transition \cite{SIDE99,ME:ECCHAPT},
hysteresis in the Br/Ag(100) system is associated with kinetic
limitations to the phase ordering/disordering processes
and critical slowing-down in the neighborhood of the second-order transition.

The model and the dynamic Monte Carlo algorithm are discussed in detail in Ref.~\cite{ME:BR2AG}
and are only briefly summarized here.
{\it In-situ} X-ray scattering indicates that the Br adlayer is commensurate
with the square Ag(100) lattice, with Br adsorbing at the four-fold hollow
sites between the Ag atoms \cite{OCKO:BR/AG}.
This arrangement suggests a lattice-gas treatment of the adlayer.
We used an $L \times L$ square lattice of adsorption sites
with periodic boundary conditions.
The grand-canonical lattice-gas Hamiltonian is \cite{ME:ECCHAPT,ME:BR2AG,KOPER:HALIDE}
\begin{equation}
{\cal H}=-\sum_{i<j} \phi_{ij} c_i c_j - \mu \sum_{i=1}^{L^2} c_i\; ,
\end{equation}
where $i$ and $j$ denote sites on the lattice,
$c_i$ denotes the occupation of site $i$
(0 for empty and 1 for occupied),
$\sum_{i<j}$ denotes a sum over all pairs of sites on the lattice, 
$\phi_{ij}$ denotes the lateral interaction energy of the pair $(i,j)$,
and $\mu$ is the electrochemical potential.
The sign convention is such that $\phi_{ij}<0$ 
denotes a repulsive interaction,
and $\mu>0$ favors adsorption.

The interactions are quite adequately 
described with a nearest-neighbor excluded-volume interaction 
(infinite repulsion) caused by the large ionic radius of Br,
plus a long-range dipole-dipole repulsion \cite{ME:BR2AG,KOPER:HALIDE}.
The lateral pair interactions are \cite{ME:BR2AG}
\begin{equation}
\phi (r)= 
\left\{
\begin{array}{ll}
-\infty & r=1 \\
2^{3/2} \phi_{\rm nnn}/r^3 & \sqrt{2} \le r \le 5\; ,\\
0 & r > 5
\end{array} 
\right.
\end{equation}
where $r$ is the separation of an interacting Br pair within the adlayer,
measured in units of the Ag(100) lattice spacing,
and $\phi_{\rm nnn}$ is the value of the 
repulsion between next-nearest neighbors.
The repulsion was truncated to simplify computation.
We found $\phi_{\rm nnn}=-26$\ meV/pair by fitting 
equilibrium simulation isotherms to experimental adsorption isotherms \cite{ME:BR2AG}.

To simulate the microscopic dynamics, 
we used a thermally activated stochastic barrier-hopping scheme,
which include adsorption, desorption, and nearest- and next-nearest neighbor diffusion.
These processes were represented by transitions 
between initial and final lattice-gas states, $I$ and $F$, respectively, 
through intermediate transition states of higher energy.
These intermediate states, $T_\lambda$, cannot be represented 
by lattice-gas configurations,
and the subscript $\lambda$ indicates the process 
(adsorption/desorption, {\it etc.}) 
which relates $I$ to $F$.
The energies of these transition states were approximated 
as \cite{ME:ECCHAPT,ME:BR2AG}
\begin{equation}
U_{T_\lambda}=(U_I+U_F)/2+\Delta_\lambda\; ,
\end{equation}
where $U_{T_\lambda}$, $U_I$, and $U_F$ are the energies 
of the states $T_\lambda$, $I$, and $F$, respectively,
and $\Delta_\lambda$ is the barrier associated with the process $\lambda$.

We approximated the probability $R(F|I)$ of making a transition 
from $I$ to $F$ during one time step 
by an Arrhenius rate \cite{ME:ECCHAPT,ME:BR2AG}
\begin{equation}
R(F|I)=\nu \exp{\left(-\frac{U_{T_\lambda}-U_I}{k_{\rm B} T}\right)}=\nu \exp{\left(-\frac{\Delta_\lambda}{k_{\rm B} T}\right)}\exp{\left(-\frac{U_F-U_I}{2 k_{\rm B} T}\right)},
\end{equation}
where $k_{\rm B}$ is Boltzmann's constant,
$T$ is the temperature,
and $\nu$ is a dimensionless attempt frequency which sets the overall 
time scale of the simulation.
For all simulations discussed here,
the temperature was 290\ K, corresponding to $k_{\rm B} T=25$\ meV.
The nearest-neighbor diffusion barrier was $\Delta_{\rm nn}=100$\ meV
and the next-nearest neighbor diffusion barrier was 
$\Delta_{\rm nnn}=200$\ meV, consistent with {\it ab initio} calculations \cite{IGNACZAK:QMHALIDE}.
The adsorption/desorption barrier was $\Delta_{\rm a}=300$\ meV,
and we used $\nu=1$ {\cite{ME:BR2AG}.
Time was measured in Monte Carlo steps per site (MCSS).

The nearest-neighbor excluded-volume interaction gives rise
to two degenerate ${\rm c}(2 \times 2)$ ordered phases,
each occupying one of two sublattices, $A$ and $B$.
The sublattice coverage, $\Theta_A$, is the fraction 
of occupied sites on sublattice $A$,
and analogously for $\Theta_B$.
There are two observables of interest:
the total Br coverage, $\Theta = (\Theta_A+\Theta_B)/2$,
and the staggered coverage $\Theta_{\rm S} = \Theta_A - \Theta_B$,
which is the ${\rm c}(2 \times 2)$ order parameter.
Figure~1 shows equilibrium $\Theta$ and $|\Theta_{\rm S}|$ isotherms
obtained by Monte Carlo simulation 
using a single-site Metropolis algorithm \cite{ME:BR2AG}.
The transition occurs at a critical electrochemical
potential, $\mu_c=180\pm 5$\ meV/particle \cite{ME:BR2AG}.

Although hysteresis studies in magnetic systems usually use
sinusoidally varying fields, CV experiments mostly use 
linearly ramped electrochemical potentials.
For the hysteresis (dynamic CV) simulations,
the Br/Ag(100) system was first equilibrated in the low-coverage phase
at $\mu_1\ll\mu_c$.
Then $\mu$ was ramped linearly in time up to  $\mu_2\gg\mu_c$
at a rate of $\rho$\ meV/MCSS.
After reaching $\mu_2$, the potential was ramped
back down to $\mu_1$ at the same rate \cite{ME:BR2AG}.
Because of the long simulation times,
only a single cycle was simulated for each value of $\rho$.

Figure~2 shows hysteresis loops, $\Theta$ vs. $\mu$,
which are analogous to magnetization vs. field plots commonly
shown for hysteresis in magnets \cite{SIDE99,X}.
However, for CVs, ${\rm d} \Theta/ {\rm d} t$ vs. $\mu$ is usually shown,
since ${\rm d} \Theta/ {\rm d} t$ is directly measurable as proportional to the electric current
through the electrochemical cell, Fig.~3.
This figure also shows a quasi-equilibrium CV,
obtained by differentiating the equilibrium coverage
isotherm in Fig.~1 with respect to $\mu$.
In the limit of zero sweep rate, the dynamic CVs should have
the same shape and $\mu$ placement as the quasi-equilibrium 
CV for the same system size.
The asymmetry between the positive-going scans (positive ${\rm d} \Theta/ {\rm d} t$)
and the negative-going scans is caused by the difference 
in the ordering (positive scan) and disordering (negative scan) dynamics.
In the limit of zero sweep rate, positive and negative scans would be symmetric.
Figure~4 shows the height, $({\rm d} \Theta/ {\rm d} t)_{\rm max}$, 
and position, $\mu_p$, of the sharp peak in the positive-going scans vs. $\rho$.

In conclusion, we have studied the hysteresis of the Br/Ag(100) adlayer
by dynamic Monte Carlo simulation.
We observed asymmetry between the positive and negative-going scans,
which is expected to vanish in the limit of zero sweep rate.
We also observed a power-law relationship between $\rho^{-1} ({\rm d} \Theta/ {\rm d} t)_{\rm max}$
and $\rho$ with a measured effective exponent of $\approx-$0.08.
Future work will include slower sweep rates to test these observed behaviors.
\\

This research was supported in part by the National Science Foundation through Grants
No.\ DMR-9634873 and DMR-9981815, and by
Florida State University through the Center for
Materials Research and Technology, 
the Supercomputer Computations Research Institute 
(US Department of Energy Contract No.\ DE-FC05-85ER25000), and 
the School of Computational Science and Information Technology.

\begin{figure}
\centerline{\epsfysize=6in \epsfbox{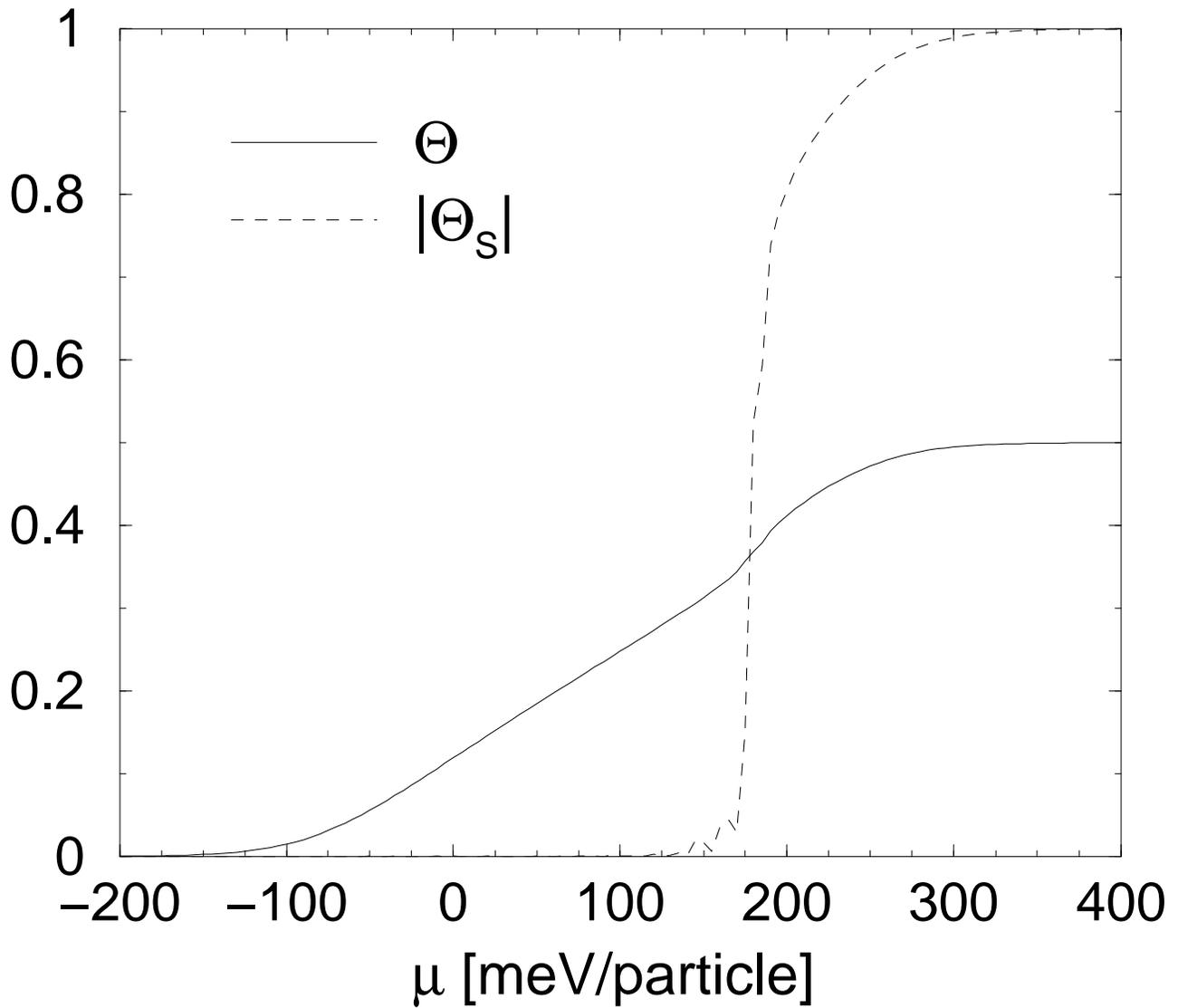}}
\caption{Equilibrium Monte Carlo isotherms for $L$=32.
A second-order phase transition between a low-coverage disordered phase and a ${\rm c}(2 \times 2)$
phase with $\Theta$=1/2 is observed at $\mu_c$.
The isotherm was generated from 10,000 independent samples for each value of $\mu$.}
\label{fig1}
\end{figure}

\begin{figure}
\centerline{\epsfysize=6in \epsfbox{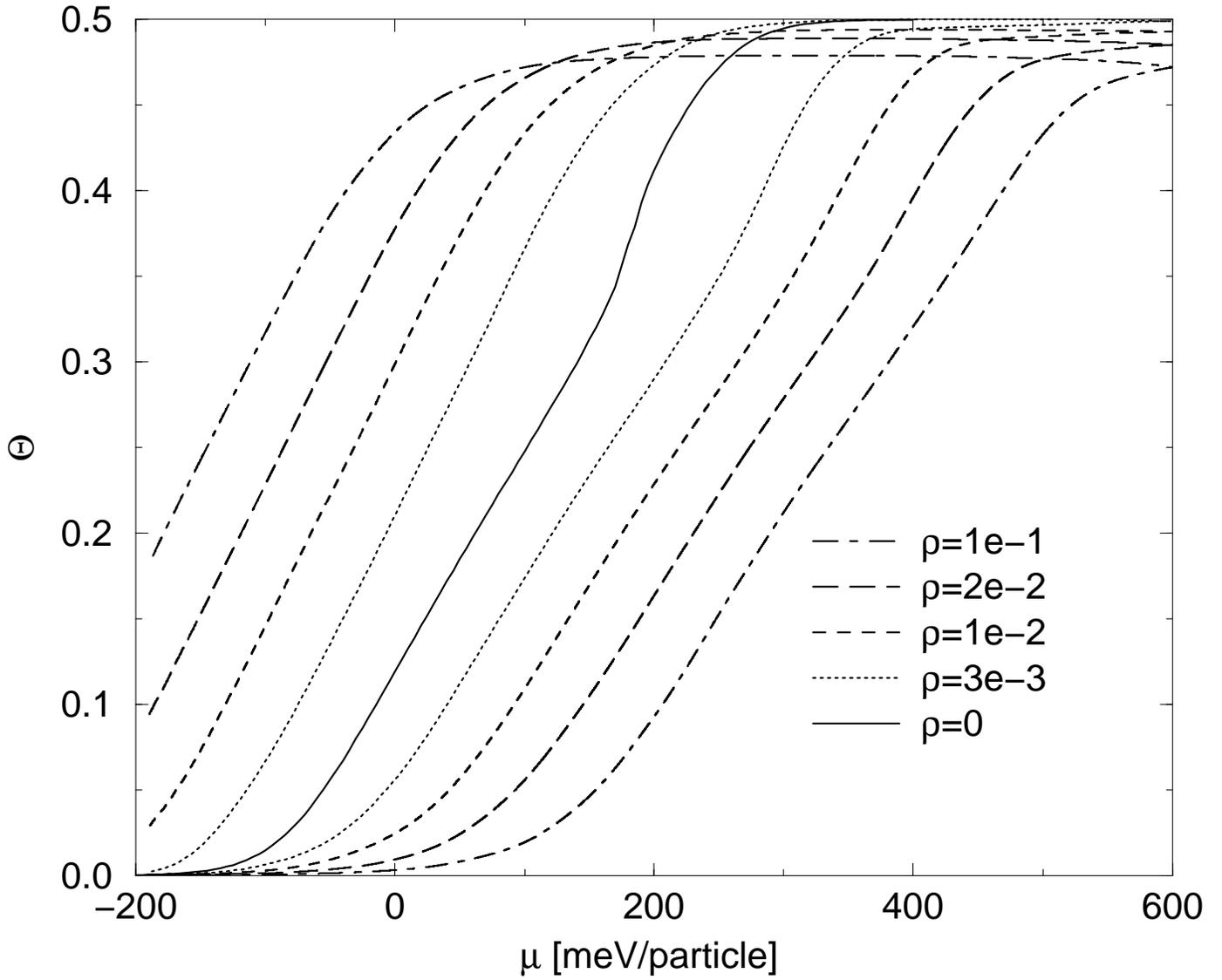}}
\caption{Hysteresis loops at various sweep rates $\rho$.
All curves show $\Theta$ vs. $\mu$ for $L$=256,
except for the solid curve, $\rho$=0, 
which shows equilibrium for $L$=32, Fig.~1.
}
\label{fig2}
\end{figure}

\begin{figure}
\centerline{\epsfysize=6in \epsfbox{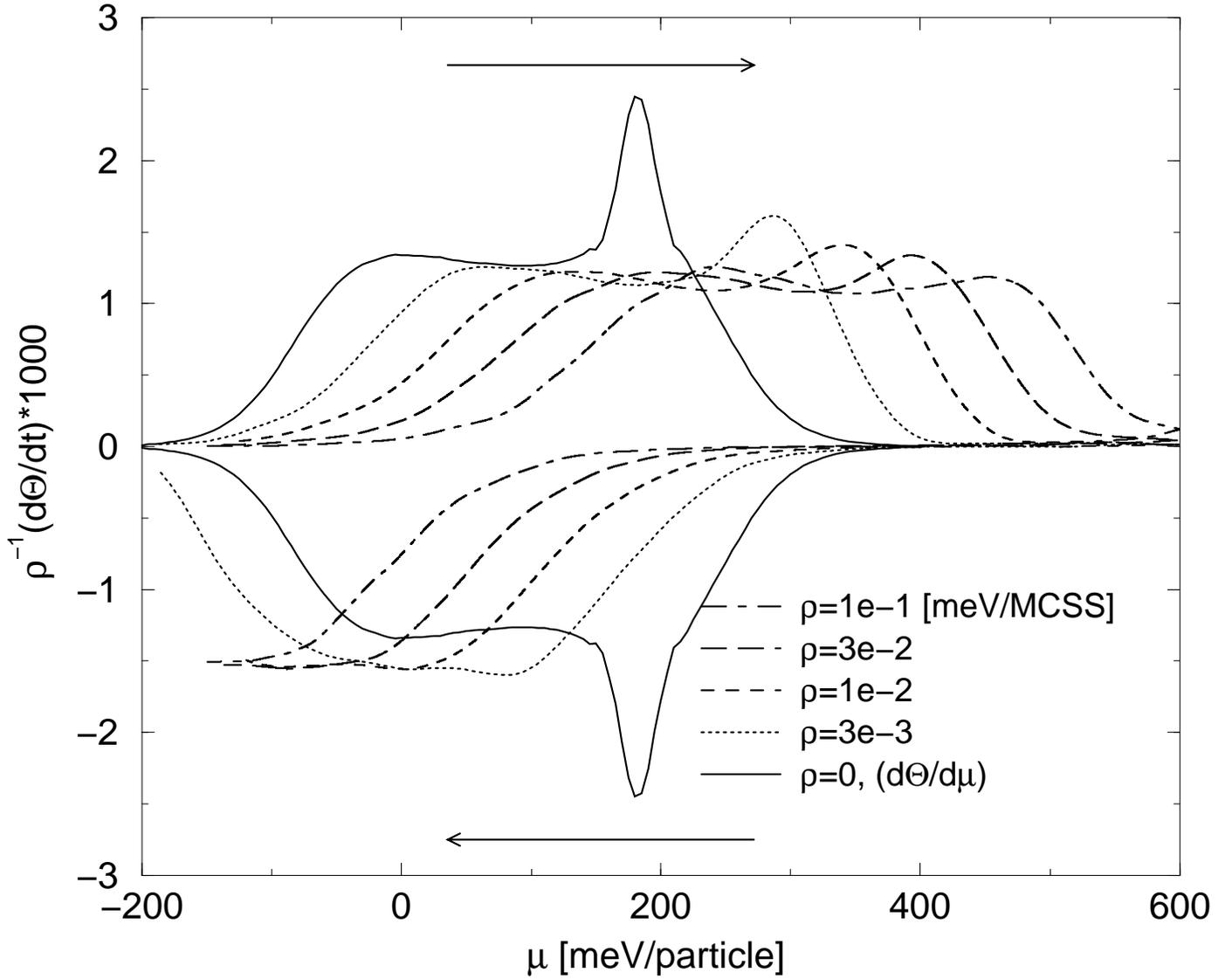}}
\caption{Simulated CVs for $L$=256 at various sweep rates $\rho$.
All curves show ${\rm d} \Theta/ {\rm d} t$ vs. $\mu$,
except for the solid curve, $\rho$=0, which shows ${\rm d} \Theta/ {\rm d} \mu$
at equilibrium for $L$=32.
All CV curves, except $\rho$=0, have been normalized by the sweep rate.
}
\label{fig3}
\end{figure}

\begin{figure}
\centerline{\epsfxsize=6in \epsfbox{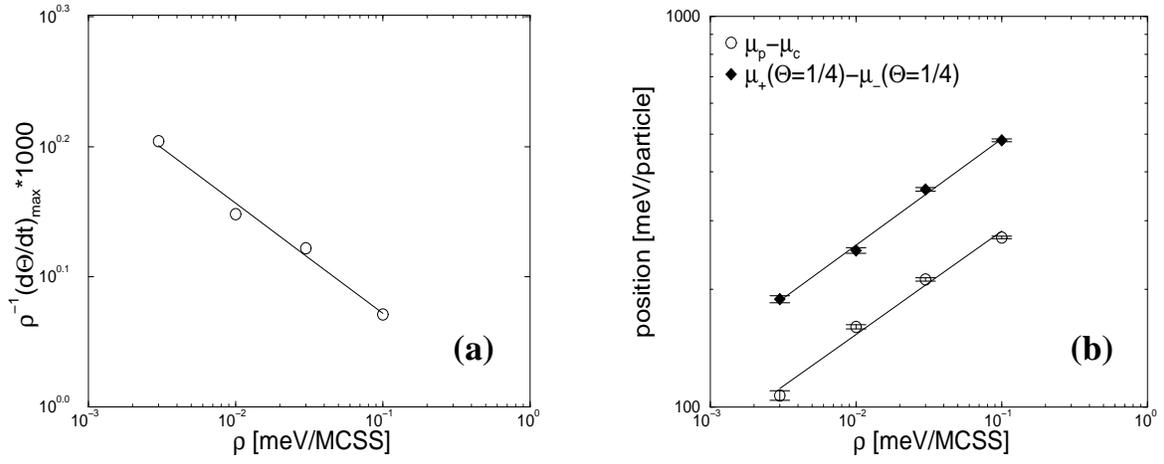}}
\caption{CV peak position and height vs. sweep rate.
{\bf (a)} shows the positive-going peak height with errors much smaller than the symbol size.
{\bf (b)} shows the peak displacement from $\mu_c$, $\mu_p-\mu_c$,
and the distance in $\mu$ between $\Theta$=1/4 
in the positive-going and negative-going scans in Fig.~2.
}
\label{fig4}
\end{figure}

\end{document}